\documentclass[lotsofwhite,charterfonts, letter]{patmorin}
\usepackage{amsmath, amsthm,amsfonts,graphicx,color,marvosym}
\usepackage[sort&compress,numbers]{natbib}

\usepackage{amsthm}

\newcommand{\Oh}[1]{\ensuremath{\protect\mathcal{O}(#1)}}
\newcommand{\sm}{\ensuremath{\setminus}}
\newcommand{\se}{\ensuremath{\subseteq}}

\newcommand{\comment}[1]{}

\newcommand{\seclabel}[1]{\label{sec:#1}}

\newcommand{\secref}[1]{\mbox{Section~\ref{sec:#1}}}

\newcommand{\figlabel}[1]{\label{fig:#1}}

\newcommand{\figref}[1]{\mbox{Figure~\ref{fig:#1}}}

\renewcommand{\eqref}[1]{(\ref{eq:#1})}

\newtheorem{thm}{Theorem}{\bfseries}{\itshape}
\newcommand{\thmlabel}[1]{\label{thm:#1}}
\newcommand{\thmref}[1]{Theorem~\ref{thm:#1}}

\newtheorem{lem}{Lemma}{\bfseries}{\itshape}
\newcommand{\lemlabel}[1]{\label{lem:#1}}
\newcommand{\lemref}[1]{Lemma~\ref{lem:#1}}

{\bfseries}{\itshape}

{\bfseries}{\itshape}

{\bfseries}{\itshape}

{\bfseries}{\itshape}

\newtheorem{assumption}{Assumption}{\bfseries}{\rm}

\usepackage[total={6.5in,9.65in}, top=1in, left=0.9in, includefoot]{geometry}

\newcommand{\sego}[2]{\ensuremath{
\raisebox{1ex}{\tiny$($\hspace*{-0.1em}}
\overline{\hspace*{0.05em}#1#2\hspace*{0.05em}}
\raisebox{1ex}{\hspace*{-0.1em}\tiny$)$}}}

\newcommand{\Figure}[4][htb]{
\begin{figure}[#1]
  \vspace*{1ex}
  \begin{center}#3\end{center}
	\vspace*{-2ex}
	\caption{\figlabel{#2}#4}
\end{figure}}

\renewcommand{\gg}{geometric graph}

\newcommand{\pg}{crossing-free geometric graph}

\newcommand{\xx}{\ensuremath{\protect{x}}}
\newcommand{\yy}{\ensuremath{\protect{y}}}
\newcommand{\zz}{\ensuremath{\protect{z}}}

\newcommand{\fl}{\ensuremath{\protect{\ell}}}
\newcommand{\fr}{\ensuremath{\protect{r}}}
\newcommand{\mv}[2][]{\ensuremath{\textup{\textsf{fix}}_{#1}(#2)}}

\newcommand{\eg}{\ensuremath{\mathcal{E}}}
\newcommand{\fg}{\ensuremath{\mathcal{F}}}
\newcommand{\hg}{\ensuremath{\mathcal{H}}}
\newcommand{\po}{\ensuremath{<_\mathcal{F}}}

\newcommand{\x}{\ensuremath{\protect\textup{\textsf{x}}}}
\newcommand{\y}{\ensuremath{\protect\textup{\textsf{y}}}}

\newcommand{\rf}[1]{\ensuremath{\textup{\textsf{roof}}(#1)}}
\newcommand{\lrf}[1]{\ensuremath{\textup{\textsf{Lroof}}(#1)}}
\newcommand{\rrf}[1]{\ensuremath{\textup{\textsf{Rroof}}(#1)}}

\renewcommand{\thefootnote}{\fnsymbol{footnote}}

\title{\MakeUppercase{A polynomial bound for untangling geometric planar graphs}}

\author{
	Prosenjit Bose\,\footnotemark[1] \and
	Vida Dujmovi\'c \,\footnotemark[2] \and
	Ferran Hurtado\,\footnotemark[3] \and
    Stefan Langerman\,\footnotemark[4]  \and
	Pat Morin\,\footnotemark[1] \and
	David R. Wood\,\footnotemark[3]
}

\footnotetext[1]{School of Computer Science, Carleton University, Ottawa, Canada. Email:\texttt{\{jit, morin\}@scs.carleton.ca}. Research partially supported by NSERC.}

\footnotetext[2]{Department of Mathematics and Statistics, McGill University, Montreal, Canada. Email:\texttt{vida@cs.mcgill.ca}. Research partially supported by CRM and NSERC.}

\footnotetext[3]{Departament de Matem\`atica Aplicada II, Universitat Polit\`ecnica de Catalunya, Barcelona, Spain. Email:\texttt{\{Ferran.Hurtado, david.wood\}@upc.edu}. Research supported by projects MEC MTM2006-01267 and DURSI 2005SGR00692. The research of David Wood is supported by a Marie Curie
Fellowship of the European Community under contract MEIF-CT-2006-023865.}

\footnotetext[4]{Chercheur Qualifi\'e du FNRS, D\'epartement d'Informatique,
Universit\'e Libre de Bruxelles, Brussels, Belgium. \texttt{Email:stefan.langerman@ulb.ac.be}.}

\date{}

\begin{document}

\maketitle

\renewcommand{\thefootnote}{\arabic{footnote}}



\begin{abstract}
To untangle a geometric graph means to move some of the vertices so that the resulting geometric graph has no crossings. Pach and Tardos [\emph{Discrete Comput.\ Geom.}, 2002] asked if every $n$-vertex geometric planar graph can be untangled while keeping at least $n^{\epsilon}$ vertices fixed. We answer this question in the affirmative with $\epsilon=1/4$. The previous best known bound was $\Omega(\sqrt{\log n / \log\log n})$. We also consider untangling geometric trees. It is known that every $n$-vertex geometric tree can be untangled while keeping at least $\sqrt{n/3}$ vertices fixed, while the best upper bound was $\Oh{(n\log n)^{2/3}}$. We answer a question of Spillner and Wolff [\texttt{http://arxiv.org/abs/0709.0170}, 2007] by closing this gap for untangling trees. In particular, we show that for infinitely many values of $n$, there is an $n$-vertex geometric tree that cannot be untangled while keeping more than $3(\sqrt{n}-1)$ vertices fixed. Moreover, we improve the lower bound to $\sqrt{n/2}$.
%
\end{abstract}
\newpage

\section{Introduction}\seclabel{intro}

Geometric reconfigurations consider the following fundamental problem. Given a starting and a final configuration of an object $\mathcal R$, determine if $\mathcal R$ can move from the starting to the final configuration, subject to some set of movement rules. An object can be a set of disks in the plane, or a graph representing a protein, or a robot's arm, for example. Typical movement rules include maintaining connectivity of the object and avoiding collisions or crossings.

In this paper we study the problem where the object is a planar graph\footnote{We consider graphs that are simple, finite, and undirected.  The vertex set of a graph $G$ is denoted by $V(G)$, and its edge set by $E(G)$. The subgraph of $G$ induced by a set of vertices $S\se V(G)$ is denoted by $G[S]$. $G\setminus S$ denotes $G[V(G)\setminus S]$.} $G$.
The starting configuration is a drawing of $G$ in the plane with vertices as distinct points and edges as straight-line segments (and possibly many crossings). 
Our goal is to relocate as few vertices of $G$ as possible in order to remove all the crossings, that is, to reconfigure $G$ to some straight line crossing-free drawing of $G$. More formally, a \emph{geometric graph} is a graph whose vertices are distinct points in the plane (not necessarily in general position) and whose edges are straight-line segments between pairs of points. If the underlying combinatorial graph of $G$ belongs to a class of graphs $\mathcal K$, then we say that $G$ is a \emph{geometric $\mathcal K$ graph}. For example, if $\mathcal K$ is the class of planar graphs, then $G$ is a geometric planar graph. Where it causes no confusion, we do not distinguish between the geometric graph and its underlying combinatorial graph. Two edges in a \gg\ \emph{cross} 
 if they intersect at some point other than a common endpoint. A \gg\ with no pair of crossing edges is called \emph{crossing-free}.

Consider a \gg\ $G$ with vertex set $V(G)=\{p_1, \dots, p_n\}$. A \pg\ $H$  with vertex set $V(H)=\{q_1, \dots, q_n\}$ is called an \emph{untangling} of $G$ if for all $i,j\in \{1,2,\dots,n\}$, $q_i$ is adjacent to $q_j$ in $H$ if and only if $p_i$ is adjacent to $p_j$ in $G$. Furthermore, if $p_i=q_i$ then we say that $p_i$ is \emph{fixed}, otherwise we say that $p_i$ is \emph{free}. If $H$ is an untangling of $G$ with $k$ vertices fixed, then we say that $G$ can be \emph{untangled} while keeping $k$ vertices fixed. Clearly only geometric planar graphs can be untangled. Moreover, since every planar graph is isomorphic to some \pg\ \cite{Wagner37, Fary48}, trivially every geometric planar graph can be untangled while keeping at least $2$ vertices fixed. For a \gg\ $G$, let \mv{G}\ denote the maximum number of vertices that can be fixed in an untangling of $G$.


At the $5$th Czech-Slovak Symposium on Combinatorics in Prague in 1998, Mamoru Watanabe asked if every geometric cycle (that is, all polygons) can be untangled while keeping at least $\varepsilon n$ vertices fixed. \citet{PT} answered that question in the negative by providing an $\Oh{(n\log n)^{2/3}}$ upper bound on the number of fixed vertices. Furthermore, they proved that every geometric cycle can be untangled while keeping at least $\sqrt{n}$ vertices fixed.

\citet{PT} asked if every geometric planar graph can be untangled while keeping $n^\varepsilon$ vertices fixed, for some $\varepsilon>0$. In recent work, \citet{sw-upg-08} showed that geometric planar graphs can be untangled while keeping $\Omega(\sqrt{\log n/\log\log n})$ vertices fixed. The best known bound before that was $3$ \cite{Goaoc}. In \secref{planar}, we answer the question of \citet{PT} in the affirmative and provide the first polynomial lower bound for untangling geometric planar graphs. Specifically, our main result is that every $n$-vertex geometric planar graph can be untangled while keeping $(n/3)^{1/4}$ vertices fixed.

There has also been considerable interest in untangling specific classes of geometric planar graphs. \citet{sw-upg-08} studied the untangling of geometric outerplanar graphs and showed that they can be untangled while keeping $\sqrt{n/3}$ vertices fixed; and that for every sufficiently large $n$, there is an $n$-vertex outerplanar graph that cannot be untangled while keeping more than $2\sqrt{n-1}-1$ vertices fixed. Thus $\Theta(\sqrt{n})$ is the tight bound for outerplanar graphs. A $\sqrt{n/3}$ lower bound for trees was shown 
 by \citet{Goaoc}. The best known upper bound for trees was \Oh{(n\log n)^{2/3}}, which was in fact proved for geometric paths, by \citet{PT}. In fact, \citet{PT} prove this upper bound for geometric cycles. However, their method readily applies for geometric paths. We answer a question posed by \citet{sw-upg-08} and close the gap for trees, by showing that for infinitely many values of $n$, there is a forest of stars that cannot be untangled while keeping more than $3(\sqrt{n}-1)$ vertices fixed. This result is proved in \secref{trees}.  In addition, in \secref{trees-lower}, we demonstrate that every geometric tree can be untangled while keeping $\sqrt{n/2}$ vertices fixed, thus slightly improving the $\sqrt{n/3}$ lower bound of \citet{Goaoc}. We conclude the paper with some open problems.


Untangling graphs has also been studied in \cite{oleg,oleg2}. \citet{Goaoc} also studied the computational complexity of the related optimization problems and showed various hardness results. 


\section{Lower bounds -- a useful lemma}\seclabel{helper}

When proving lower bounds, our goal will be to show that given any geometric planar graph $G$ we can find a large subset $R$ of vertices of $G$ such that $G$ can be untangled while keeping $R$ fixed. The following geometric lemma simplifies this task by allowing us to concentrate on the case in which all vertices of $R$ are on the \y-axis. This lemma will be useful both for untangling geometric trees in \secref{trees-lower} and for untangling general geometric planar graphs in \secref{planar}.

\begin{lem}\lemlabel{simplify}
Let $\overline{G}$ be an untangling of some geometric planar graph $G$. 
Let $R$ be a set of vertices of $G$ such that each vertex of $R$ is on the \y-axis in $\overline{G}$ and has the same \y-coordinate in $\overline{G}$ as in $G$. Then there
exists an untangling $\overline{G'}$ of $G$ in which the vertices in
$R$ are fixed.
\end{lem}

\begin{proof}
The proof uses the fact that it is possible to perturb the vertices of a \pg\ without introducing crossings. More precisely, for any \pg\ there exists a value $\varepsilon>0$ such that each vertex can be moved a distance of at most $\varepsilon$, and the resulting geometric graph is also crossing-free. The maximum value $\varepsilon$ for which this property holds is called the \emph{tolerance} of the arrangement of segments. This concept, both for the geometric realization and the combinatorial meaning of the graphs was systematically studied in \cite{Ramos-phd,ferran-pedro}.


Consider the untangling $\overline{G}$ of $G$ and let $\varepsilon>0$ be the value obtained when the above perturbation fact is considered for to $\overline{G}$. Let $X$ denote the maximum absolute value of an \x-coordinate in $G$ of a vertex in $R$. Let $\overline{G''}$ be the geometric graph obtained from $\overline{G}$ as follows. For each vertex $v\in R$ positioned at $(\x,\y)$ in $G$, move $v$ from $(0,\y)$ in $\overline{G}$ to $(\x\varepsilon/X, \y)$ in $\overline{G''}$. The vertices not in $R$ are unmoved. So each vertex moves a distance of at most $\varepsilon$, and $\overline{G''}$ is crossing-free. Scale $\overline{G''}$ by multiplying the \x-coordinates of all vertices in $\overline{G''}$ by $X/\varepsilon$ to obtain a \pg\ $\overline{G'}$. Then every vertex of $R$ has the same location in  $\overline{G'}$ as it does in $G$. Thus $\overline{G'}$ is an untangling of $G$ that keeps the vertices of $R$ fixed.
\end{proof}

\section{Trees -- lower bound}\seclabel{trees-lower}
%
%

\citet{Goaoc} proved a lower bound of $\mv{T} \geq \sqrt{n/3}$ for  every $n$-vertex geometric tree $T$. We now give a different construction that yields a slightly better constant. In addition to an improved constant, our motivation for including this result is that it provides a warm up to our main result, the polynomial lower bound for planar graphs.

\begin{thm}\thmlabel{trees-lower}
Every $n$-vertex geometric tree $T$ can be untangled while keeping at least $\sqrt{n/2}$ vertices fixed. That is, $\mv{T} \geq \sqrt{n/2}$. 
\end{thm}

In a vertex $2$-colouring of $T$, the largest of the two colour classes has at least $n/2$ vertices. Therefore, the following lemma, coupled with \lemref{simplify}, implies \thmref{trees-lower}. 

\begin{lem}\lemlabel{tree-helper}
Let $T$ be an $n$-vertex geometric tree $T$ whose vertices are $2$-coloured. Let $S$ be one of the two colour classes. Then there exists a set $R$ of vertices in $T$ such that $|R|\geq\sqrt{|S|}$ and there is an untangling $T'$ of $T$ in which each vertex in $R$ 
is on the \y-axis and has the same \y-coordinate in $T'$ as in $T$.
\end{lem}

\begin{proof}
Root $T$ at any vertex and label its vertices $(v_1,\dots, v_n)$ based on a postorder traversal of $T$. 

While we make no general position assumption on the vertices of $T$, we may assume, by a suitable rotation, that no pair of vertices of $T$ have the same \y-coordinate.  Let $R$ be a largest ordered subset $R\subseteq S$ such that the \y-coordinates of the vertices of $R$ are either monotonically increasing or monotonically decreasing when considered in the order $(v_1,\dots, v_n)$. By the Erd{\H{o}}s-Szekeres Theorem \cite{ES35}, $|R|\geq \sqrt{|S|}$. Without loss of generality, assume $R$ is monotonically increasing. 

Let $T'$ be a geometric tree obtained from $T$ as follows.  For each vertex $v\in R$ positioned at $(x, y)$ in $T$, move $v$ to $(0,y)$ in $T'$. Move each vertex in $S\sm R$ from its position in $T$ to the \y-axis, such that all the vertices of $S$ in $T'$ appear in the order $\sigma$ on the \y-axis. The vertices in $V(T)\sm S$ remain unmoved. To complete the proof of the lemma, it remains to show how to untangle $T'$ while keeping $S$ fixed. We prove that by induction on $i$, with the following induction hypothesis. 

For a point $p$ with coordinates $(x,y)$, a \emph{right ray} at $p$ is the open half-line containing all the points $(x',y)$ where $x'>x$. Let $T_i:= T'[\{v_1, \dots, v_i\}]$.
For each $i \in \{1,\dots,n\}$, there is an untangling $\overline{T_i}$ of  $T_i$ such that $S \cap V(T_i)$ is fixed, and\\~\\
$(1)$ for all $j\in\{2,\dots,i\}$, the \y-coordinate of $v_j$ is greater than the \y-coordinate of $v_{j-1}$, and\\
$(2)$ for each vertex $v$ whose parent in $T$ is not in $\{v_1, \dots, v_i\}$, the right ray at $v$ does not intersect $\overline{T_i}$. \\~\\
\figref{binary} depicts such an untangling of the complete binary tree of depth $4$.
\Figure{binary}{\includegraphics[width=4in]{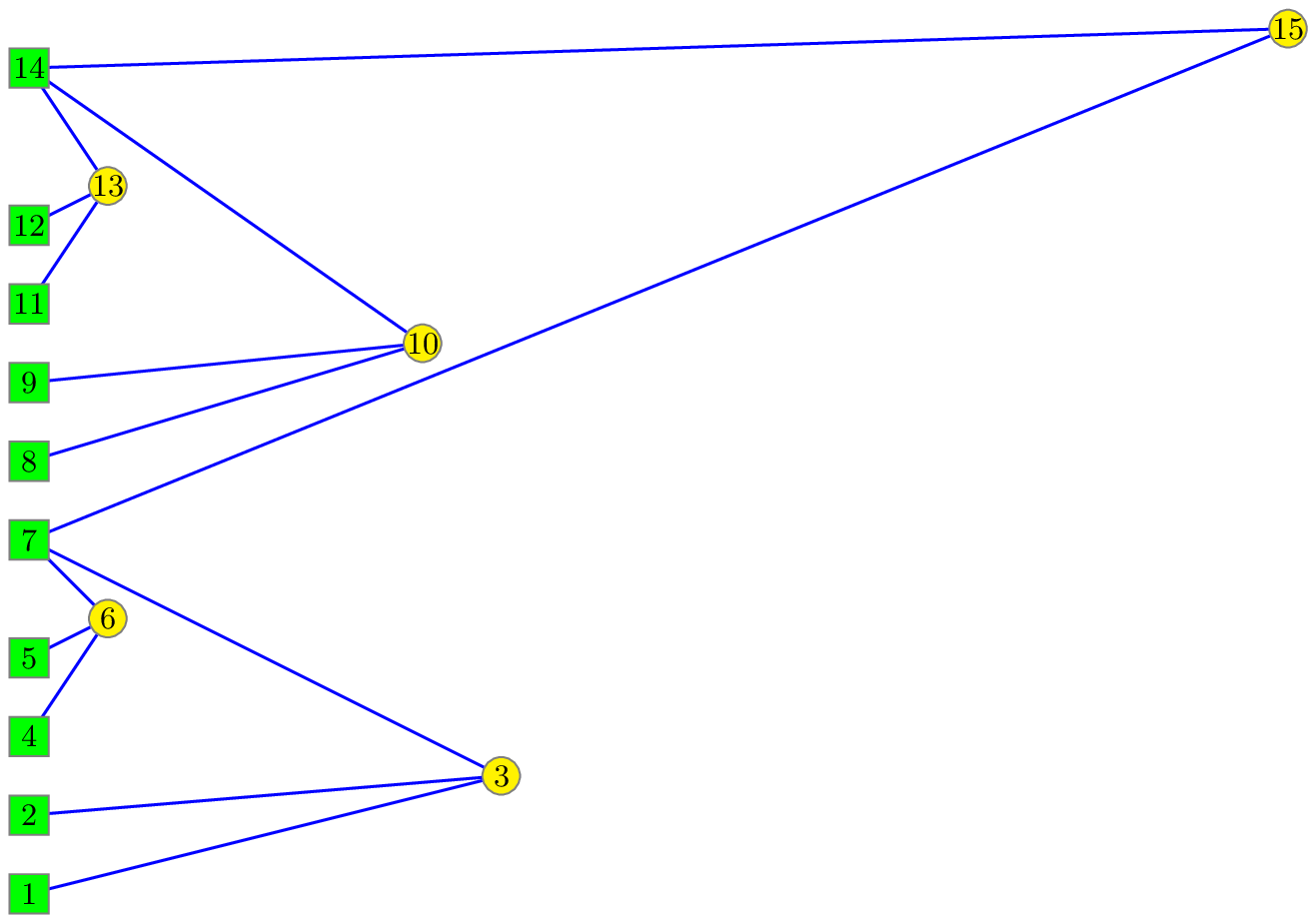}
}{An untangling of the complete binary tree of depth $4$. Vertices of $S$ are depicted by squares.}

For $i=1$, the statement is true trivially. Assume now that $i>1$, and that the statement is true for $i-1$. There are two cases to consider: $v_i\in S$ and $v_i\not\in S$. 

Consider first the case that $v_i\not\in S$. Since $S$ is a colour class in a $2$-colouring of $T$, each child of $v_i$, if any, is in $S$ and thus is on the \y-axis. Assign an \y-coordinate to $v_i$ that is greater than the \y-coordinate of each vertex in $\overline{T_{i-1}}$ and less than the \y-coordinate of each vertex in $S\sm V(T_{i-1})$. This ensures that condition $(1)$ is maintained in $\overline{T_i}$. Assign a positive \x-coordinate to $v_i$ such that $\overline{T_i}$ is crossing-free. Condition $(2)$ on $T_{i-1}$ guarantees that this is always possible. Condition $(2)$ is clearly maintained for $v_i$ in $\overline{T_i}$. The only other vertices of $\overline{T_i}$ which may violate condition $(2)$, are vertices whose \y-coordinates in $\overline{T_i}$ are between that of $v_i$ and its smallest indexed child. However, since the vertices are indexed by the postorder traversal of $T$, all such vertices are in the subtree of $T$ rooted at $v_i$, and thus each of their parents is in $T_i$. Therefore condition $(2)$ is maintained in $\overline{T_i}$.

To complete the proof we consider the case that $v_i\in S$. We start with an observation.
Consider a vertex $v\in V(\overline{T_{i-1}})\setminus S$ whose parent is not in $\overline{T_{i-1}}$. Let the coordinates of $v$ in $\overline{T_{i-1}}$ be $(x,y)$. 
Each child of $v$ is in $S$ and thus lies on the \y-axis. Denote their \y-coordinates by $y_1, \dots, y_d$. By condition $(1)$, for each $i\in \{1,\dots,d\}$, the right ray at $(0,y_i)$ can only be intersected by an edge incident to $v$ in $\overline{T_{i-1}}$. Thus $v$ can be moved to any position $(x',y)$, $x'\geq 0$, and the resulting untangling of $T_{i-1}$ still satisfies the two conditions.  We are now ready to untangle $T_i$. Vertex $v_i$ is fixed, and thus its position in $\overline{T_i}$ is predetermined. None of its children are in $S$. Thus we are allowed to move any child of $v_i$ from its position in $\overline{T_{i-1}}$ to a new position. By the above observation it is possible to move each child $w$ of $v_i$ (one by one, in the decreasing order of their \y-coordinates), such that the resulting untangling  $\overline{T_{i-1}'}$ of $T_{i-1}$ satisfies conditions $(1)$ and $(2)$, and such that the open segment $\sego{w}{v_i}$ does not intersect $\overline{T_{i-1}'}$. Connect $v_i$ by a segment to each of its children in $\overline{T_{i-1}'}$. Then the resulting untangling $\overline{T_i}$ is crossing-free. Condition $(1)$ is maintained since all the vertices of $T_{i-1}$ have smaller \y-coordinate that $v_i$ in $\overline{T_i}$. Condition $(2)$ is maintained in $\overline{T_i}$ by the same arguments used when $v_i\not\in S$.
\end{proof}


\section{Planar graphs - lower bound}\seclabel{planar}

%
Let $G$ be an $n$-vertex geometric planar graph. In this section we prove that $G$ can be untangled while keeping $(n/3)^{1/4}$ vertices fixed (as stated in \thmref{planar} below). It suffices to prove this theorem for edge-maximal geometric planar graphs. Thus for the remainder of this section assume that $G$ is edge-maximal.\footnote{A planar graph $H$ is edge-maximal (also called, a \emph{triangulation}), if for all $vw\not\in E(H)$, the graph resulting from adding $vw$ to $H$ is not planar.} 

Let \eg\ be an embedded planar graph isomorphic to $G$.
 Each face of \eg\ is bounded by a $3$-cycle. 
Canonical orderings of embedded edge-maximal planar graphs were introduced by
\citet{dFPP90}, where they proved that
\eg\ has a vertex ordering $\sigma=(v_1:=\xx,v_2:=\yy, v_3, \dots,
v_n:=\zz)$, called a \emph{canonical ordering}, with the following
properties. Define $G_i$ to be the embedded subgraph of \eg\ induced by
$\{v_1, v_2,\dots,v_i\}$.  Let $C_i$ be the subgraph of \eg\ induced
by the edges on the boundary of the outer face of $G_i$. Then
\begin{itemize}
\item \xx, \yy\ and \zz\ are the vertices on the outer face of \eg, and 
\item  For each $i\in\{3,4,\dots,n\}$, $C_i$ is a cycle containing $\xx\yy$.
\item  For each $i\in\{3,4,\dots,n\}$, $G_i$ is biconnected and \emph{internally $3$-connected}; that is, removing any two interior vertices of $G_i$ does not disconnect it.
\item For each $i\in\{3,4,\dots,n\}$, $v_i$ is a vertex of $C_i$ with at least two neighbours in $C_{i-1}$, and these neighbours are consecutive on $C_{i-1}$.
\end{itemize}
For example, the ordering in \figref{canonical}(a) is a canonical ordering of the depicted embedded graph \eg.
\Figure{canonical}{\includegraphics[width=6.5in]{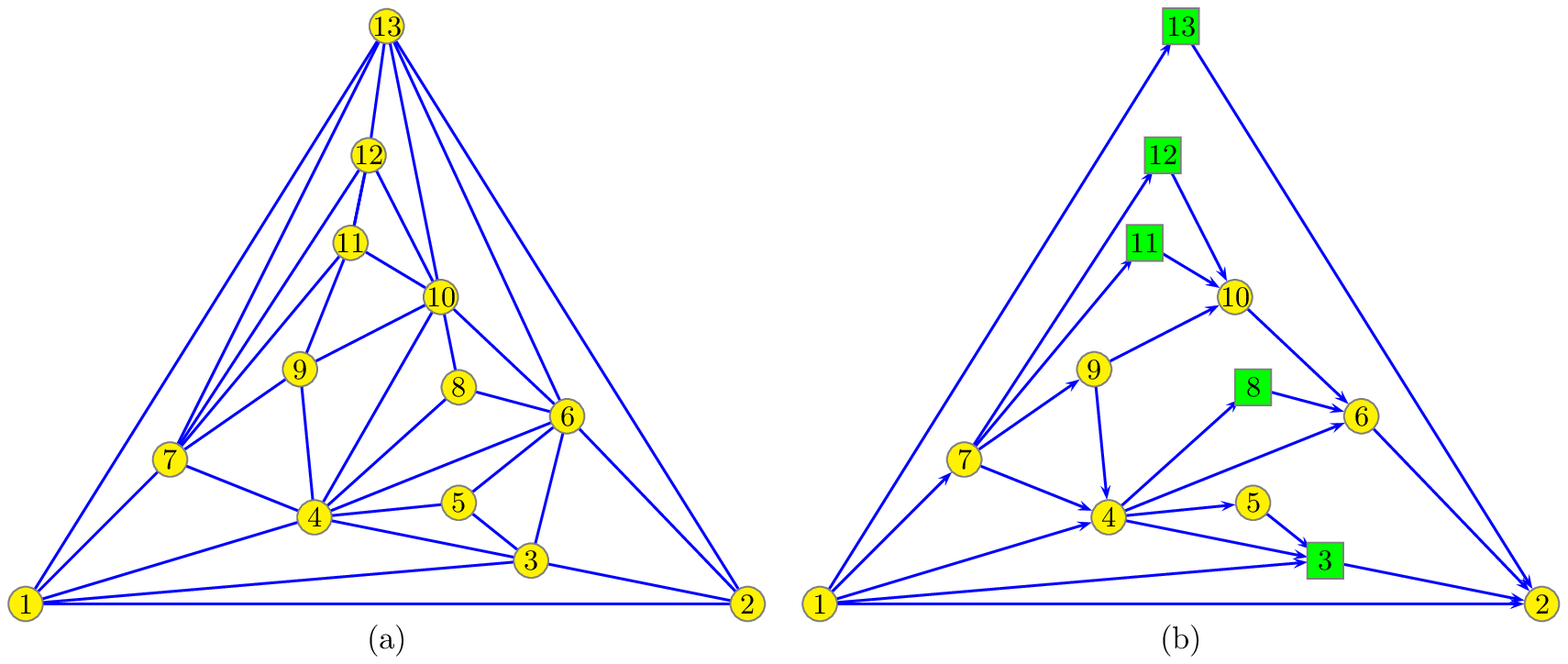}
}{ (a) Canonical ordering of \eg, (b) Frame \fg\ of \eg. Vertices forming a largest antichain in $<_\fg$, that is the vertices in $S$, are depicted by squares.}

We now introduce a new combinatorial structure that is critical to the proof of \thmref{planar}.  The \emph{frame} \fg\ of \eg\ is the oriented subgraph of \eg\ with vertex set  $V(\fg):=V(\eg)$, where: 
\begin{itemize}
\item \xx\yy\ is in $E(\fg)$ and is oriented from \xx\ to \yy.
\item For each $i\in\{3,4,\dots,n\}$ in the canonical ordering $\sigma$ of \eg, edges $pv_i$ and $v_ip'$ are in $E(\fg)$, where $p$ and $p'$ are the first and the last neighbour, respectively, of $v_i$ along the path in $C_{i-1}$ from \xx\ to \yy\ not containing edge \xx\yy. Edge $pv_i$ is oriented from $p$ to $v_i$, and edge $v_ip'$ is oriented from $v_i$ to $p'$, as illustrated in \figref{canonical}(b). We call $p$ the \emph{left predecessor} of $v$ and $p'$ the \emph{right predecessor} of $v$.
\end{itemize}

We also say that \fg\ is a frame of $G$. By definition, \fg\ is a directed acyclic graph with one source \xx\, and one sink \yy. \fg\ defines a partial order \po\ on $V(\fg)$, where $v\po w$ whenever there is a directed path from $v$ to $w$ in \fg. 

The remainder of this section is dedicated to proving the following two lemmas, which readily imply the desired result, as shown in the proof of \thmref{planar} below.

\begin{lem}\lemlabel{chain}
Every $n$-vertex geometric planar graph $G$ whose partial order \po\ associated with its frame \fg\ has a chain of size $\ell$ can be untangled while keeping $\sqrt{\ell/3}$ vertices fixed.
\end{lem}

\begin{lem}\lemlabel{antichain}
Every $n$-vertex geometric planar graph $G$  whose partial order \po\ associated with its frame \fg\ has an antichain of size $t$ can be untangled while keeping $\sqrt{t}$ vertices fixed.
\end{lem}

\begin{thm}\thmlabel{planar}
Every $n$-vertex geometric planar graph $G$ can be untangled while keeping at least $(n/3)^{1/4}$ vertices fixed. That is, $\mv{G} \geq (n/3)^{1/4}$. 
\end{thm}

\begin{proof}
Let \fg\ be a frame of $G$ and let \po\ be its associated partial order. If \po\ has a chain of size at least $\sqrt{3n}$ then we are done by \lemref{chain}. Otherwise, by Dilworth's theorem \cite{Dilworth50}, \po\ has a partition into $\sqrt{3n}$ antichains. By the pigeon-hole principle there is an antichain in that partition that has at least $\frac{n}{\sqrt{3n}}$ vertices, which completes the proof, by \lemref{antichain}.
\end{proof}

The remainder of this section is dedicated to proving \lemref{chain} and \lemref{antichain}.

\subsection{Big chain - Proof of \lemref{chain}}\seclabel{bigchain}

A \emph{chord} of a cycle $C$ is an edge  that has both endpoints in $C$, but itself is not an edge of $C$. Consider a cycle $C$ in an embedded planar graph $\mathcal E$. $C$ is called \emph{externally chordless} if each chord of $C$ is embedded inside of $C$ in $\mathcal E$. The following theorem is by \citet{sw-upg-08}.

\begin{thm}\thmlabel{alex}\cite{sw-upg-08}
Let $G$ be a geometric planar graph and \eg\ an embedding planar graph isomorphic to $G$. If \eg\ has an externally chordless cycle on $\ell$ vertices, then $G$ can be untangled while keeping at least $\sqrt{\ell/3}$ vertices fixed. Note that this result is expressed in slightly different form in \cite{sw-upg-08} (see Theorem $2$ in \cite{sw-upg-08}). 
\end{thm}

\begin{lem}\lemlabel{weakly}
Consider any directed path on at least three vertices from \xx\ to \yy\ in \fg. The cycle comprised of that path and edge $\xx\yy$ is externally chordless in \eg.
\end{lem}
\begin{proof}
Denote the cycle in question by $C$, and denote the directed path
between \xx\ and \yy\ in $C$ not containing edge \xx\yy\ by $P$. Consider a chord $v_iv_j$ of $C$. Without loss of generality, $v_i<_\sigma v_j$ in the canonical ordering $\sigma$.
Thus $v_i$ is in $G_{j-1}$ and $v_iv_j$ is an edge of $G_j$.
The neighbours of $v_j$ in $G_{j-1}$ appear consecutively along the boundary $C_{j-1}$ of $G_{j-1}$. Let $x_1,\dots, x_d$ be the neighbours of $v_j$ in left-to-right order on $C_{j-1}$. Thus $x_1v_j$ and $v_jx_d$ are arcs in \fg. Let $uv_j$ and $v_jw$ be the incoming and outgoing arcs in $P$ at $v_j$. Then the counterclockwise order of edges incident to $v_j$ in \eg\ is $(u,\dots,x_1,\dots,x_d,\dots,w,\dots)$. In particular, each edge $v_jx_\ell$ is contained in the closure of the interior of $C$. Now $v_i = x_\ell$ for some $\ell\in[1,d]$. Thus $v_iv_j$ is an internal chord of $C$.
\end{proof}

This lemma, coupled with \thmref{alex}, implies \lemref{chain}, as demonstrated below.

\begin{proof}[Proof of \lemref{chain}.]
If $\ell<3$, the claim follows trivially. Assume now that $\ell\geq
3$. Since \po\ has a chain of size $\ell$, \po\ has a maximal chain of
size $\ell'\geq \ell$. Every maximal chain in \po, is a path from \xx\
to \yy\ in \fg. Therefore, \lemref{weakly} implies that \eg\ contains
an externally chordless cycle on $\ell'$ vertices, and the result follows
from \thmref{alex}.
\end{proof}

\subsection{Big Antichain - Proof of \lemref{antichain}}\seclabel{bigantichain}
For each vertex $v\in V(\fg)$, we define  \lrf{v} and \rrf{v}, as the following directed paths in \fg. \\
$~~~~~~~~~~~~$ $\lrf{v_1}:=\emptyset$ and $\rrf{v_1}:=\emptyset$,\\
$~~~~~~~~~~~~$ $\lrf{v_2}:=\emptyset$ and $\rrf{v_2}:=\emptyset$.\\
For each $i\in \{3,\dots, n\}$, define \lrf{v_i}\ and \rrf{v_i} recursively, as follows.\\
$~~~~~~~~~~~~$ $\lrf{v_i}:=\lrf{p}\cup \{pv_i\}$, and \\
$~~~~~~~~~~~~$ $\rrf{v_i}:=\{v_ip'\}\cup \rrf{p}$, \\
where $p$ is the left and $p'$ the right predecessor of $v_i$. Finally, define the \emph{roof} of $v_i$ to be $\rf{v_i}:=\lrf{v_i}\cup\rrf{v_i}$.

Note that for each $i\in \{3,\dots, n\}$, \rf{v_i}\ is a directed path in \fg\ from \xx\ to \yy\ containing $v_i$, where the sub-path ending at $v_i$ is \lrf{v_i}, and the sub-path starting $v_i$ is \rrf{v_i}.

Let $S$ be the set of vertices that comprise a largest antichain in $<_{\mathcal F}$, as illustrated in \figref{canonical}(b) with squares. Now consider the given geometric graph $G$. We may assume, by a suitable rotation, that no pair of vertices of $G$ have the same \y-coordinate.  Let $R$ be a largest ordered subset $R\subseteq S$ such that the \y-coordinates of the vertices of $R$ are either monotonically increasing or monotonically decreasing when considered in the order given by $\sigma$. By the Erd{\H{o}}s-Szekeres Theorem \cite{ES35}, $|R|\geq \sqrt{|S|}$. Without loss of generality, assume $R$ is monotonically increasing. In what follows, we untangle $G$ while keeping $R$ fixed.

Let \hg\ be the graph induced in \eg\ by the following set of vertices: $V(\hg):= \cup\{ \rf{w}: w \in R \}$
; that is, $\hg=\eg[V(\hg)]$.  Note that \hg\ is not necessarily a subgraph of \fg, as illustrated in \figref{hg}.
\Figure{hg}{\includegraphics[width=3.5in]{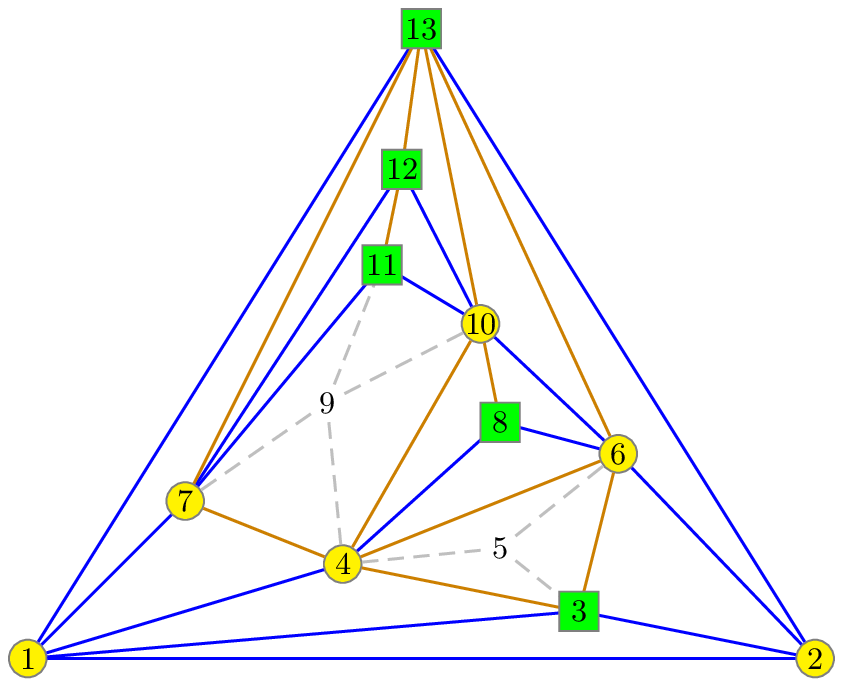}
}{The graph \hg. The vertices in $R\subseteq S$ are depicted by squares. Edges $\{ 3,6\}$, $\{4,10\}$, $\{8,10\}$, $\{6,13\}$, $\{7,13\}$, $\{10, 13\}$ and $\{12,13\}$ are in \hg\ but not in \fg.}

We say that a simple polygonal chain $C$ is \emph{strictly \x-monotone} if, for every vertical line $\ell$, $|C\cap\ell| \le 1$.
For two distinct points $p$ and $q$ in the plane, let \sego{p}{q} denote the {\em open} line-segment with endpoints $p$ and $q$. A simple polygon $C$ is \emph{star-shaped} (from $p$) if there is a point $p$ such that for every point $q\in C$, $\sego{p}{q}\cap C =\emptyset$. The following lemma is the main ingredient in the proof of \lemref{antichain}. 

\begin{lem}\lemlabel{drawing}
The geometric planar graph $G[V(\hg)]$ can be untangled such that each vertex of $R$ is on the
\y-axis and it has the same \y-coordinate in the untangling as in $G[V(\hg)]$. Moreover, all the internal faces of the untangling are star-shaped and the path on its outer face from \xx\ to \yy\ not containing \xx\yy\ is strictly \x-monotone.
\end{lem}

We delay the proof of \lemref{drawing} until the end of the section.
We first show how it implies our desired result when coupled with the
following theorem by \citet{DBLP:conf/wg/HongN06}.

\begin{thm}\cite{DBLP:conf/wg/HongN06}\thmlabel{star}
Consider a $3$-connected embedded planar graph ${\mathcal E}$, with outer facial cycle $C$. Given any geometric cycle $\overline{C}$ that is star-shaped, and given any isomorphic mapping from $V(C)$ to $V(\overline{C})$, there is a crossing-free geometric graph $\overline{\mathcal E}$ isomorphic to ${\mathcal E}$ with $\overline{C}$ as its outer face and respecting the vertex mapping.
\end{thm}

\begin{proof}[Proof of \lemref{antichain}.]
Since $<_{\mathcal F}$ has an antichain of size $t$, $<_{\mathcal F}$ has a maximal antichain $S$ of size $t'\geq t$. Then the subset $R\subseteq S$, on which \hg\ is defined, has size $|R|\geq \sqrt{t}$. Thus by \lemref{drawing}, $G[V(\hg)]$ can be
untangled such that the vertices of $R$ are all on the \y-axis and their
\y-coordinates are preserved.  If $\zz\not\in R$, then assign \x- and
\y-coordinates to \zz, and connect \zz\ to its neighbours in \hg, such
that the resulting geometric graph $H$ is crossing-free and all the internal
faces of $H$ are star-shaped. This is always possible since the path from \xx\ to \yy\ on the outer face of the above untangled graph is strictly \x-monotone. $H$ is an untangling of $G[V(\hg)\cup \{\zz\}]$.


It remains to determine a placement of the remaining free vertices of $G$, that is vertices in $V(G)\sm V(H)$.
Vertices of $V(G)\sm V(H)$ can be partitioned into sets $I_j$, $1
\leq j\leq |E(H)|-|V(H)|+1$, where each vertex in $I_j$ is inside the cycle in \eg\ determined by the internal face $f_j$ of $H$. For
each internal face $f_j$ of $H$, let $G^j$ be the following subgraph
of \eg. The vertex set $V(G^j)$ is the union of $V(f_j)$ and $I_j$. The edge set $E(G^j)$ is comprised of the edges of the cycle $f_j$, the edges in $\eg[I_j]$, and the edges between $V(f_j)$ and $I_j$. Each $f_j$ is star-shaped in
$H$, by \lemref{drawing}. Therefore, to apply \thmref{star}, it
remains to show that $G^j$ is $3$-connected. 

Assume, for the sake of contradiction, that $G^j$ is not
$3$-connected. All the faces of $G^j$ are triangles except possibly
the outer face $C^j$. Therefore, $G^j$ is internally $3$-connected,
that is, removing any two interior vertices of $G^j$ does not
disconnect it. Thus each cut-set of size $2$ of $G^j$ has a vertex,
say $v$, that is in $C^j$.  Removing $v$ from $G^j$ results in a
graph that is not $2$-connected. The outer face $C^j$ has no chords,
since $f_j$ is a face of $H$. Therefore, removing $v$ from $G^j$
results in graph whose outer face is a cycle and all internal faces
are triangles. Thus that graph is a $2$-connected graph, which provides
the contradiction.  

Applying \thmref{star} to embed each subgraph $G^j$ yields an
untangling of $G$ in which the vertices of $R$ are all on the \y-axis
and have their \y-coordinates preserved.  Applying \lemref{simplify}
to this untangling completes the proof of the theorem.
\end{proof}

All that remains is to prove \lemref{drawing}.


\begin{proof}[Proof of \lemref{drawing}]
The proof is by induction on the number of vertices in $R$. We start by considering some useful properties of the roofs of two vertices in $R$.

Consider two incomparable vertices, $u$ and $v$ in $R$ (that is, two incomparable vertices in $<_\fg$), where $u<_\sigma v$. Let $\xx'$ be a vertex of \fg\ such that $\xx'\in\lrf{u}$ and $\xx'\in\lrf{v}$, and the vertex following $\xx'$ in $\lrf{u}$ is not the same as the vertex following $\xx'$ in $\lrf{v}$, as illustrated in \figref{incomparable}.  Similarly, let $\yy'$ be a vertex of \fg\ such that $\yy'\in\rrf{u}$ and $\yy'\in\rrf{v}$, and the vertex before $\yy'$ in $\rrf{u}$ is not the same as the vertex before $\yy'$ in $\rrf{v}$. Such vertices, $\xx'$ and $\yy'$, exist since $u$ and $v$ are incomparable in \fg. Then \rf{v} and \rf{u} have the following properties. The paths between \xx\ and $\xx'$ in \rf{u}\ and in \rf{v}\ coincide in \fg, that is, the two paths are both equal to \lrf{\xx'}. Similarly, the paths between $\yy'$ and $\yy$ in \rf{u}\ and in \rf{v}\ coincide in \fg, that is, they are both equal to $\rrf{\yy'}$. The path between $\xx'$ and $\yy'$ in \rf{u} contains $u$, the path between $\xx'$ and $\yy'$ in \rf{v} contains $v$, and the two paths have only $\xx'$ and $\yy'$ in common. Finally, $u$ is inside the cycle determined by \rf{v}\ and edge \xx\yy\ in \fg. To summarise, for all $u,v\in R$, if $u<_\sigma v$ then each vertex of \rf{u}\ is either on or inside the cycle determined by \rf{v}\ and edge \xx\yy\ in \fg. 
\Figure{incomparable}{\includegraphics[width=3in]{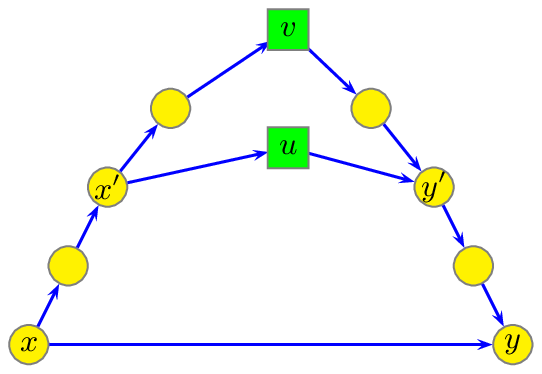}                
}{Roofs of two incomparable vertices $u$ and $v$ of $<_\fg$.}

We proceed by induction on the number of vertices in $R$, but require a somewhat stronger inductive hypothesis than the statement of the lemma.  Let $C$ be a simple strictly \x-monotone polygonal chain.  We say that $C$ is \emph{$\varepsilon$-ray-monotone} from a point $p=(x_p,y_p)$ if for every point $r=(x_p,y_p+t)$ with $t\ge\varepsilon$, and every point $q\in C$, $\sego{r}{q}\cap C=\emptyset$.  Informally, $C$ is $\varepsilon$-ray-monotone from $p$ if every point sufficiently far above $p$ sees all of $C$. Note that, under this definition, if $C$ is $\varepsilon$-ray-monotone from $p$ then $C$ is $\varepsilon$-ray-monotone from any point $q=(x_p,y_p+t)$, $t>0$, above $p$. Furthermore, there exists a value $\delta=\delta(p,C,\varepsilon)$ such that $C$ is $\varepsilon$-ray-monotone from any point $p'$ whose distance from $p$ is at most $\delta$.  (This follows from the fact that the set of points $p$ from which $C$ is $\varepsilon$-ray-monotone is an open set.)

Let $\varepsilon'$ be the minimum difference between the \y-coordinates of
any two vertices in $R$.  We will construct a \pg\ 
$\overline{\hg}$ that is an untangling of $G[V(\hg)]$. In addition to the conditions of the lemma, $\overline{\hg}$ will have the following property:
If $|R|>0$ then the outer face of $\overline{\hg}$ is bounded by the edge
$xy$ and a path $C$ from $x$ to $y$ such that $C\cap R=\{v\}$, for some vertex $v\in R$, and $C$ is $\varepsilon$-ray-monotone from $v$ for some $\varepsilon < \varepsilon'$.

The base case occurs when $|R|=0$. Then $\hg$ consists of the single edge \xx\yy, which can be untangled by placing \xx\ at $(-1,t)$ and \yy\ at $(1,t)$, where $t$ is smaller than any \y-coordinate in $G$.  Clearly this \pg\ satisfies the conditions of the lemma
as well as the inductive hypothesis.  Next, suppose $|R|\ge 1$ and let
$v$ be the largest vertex of $R$ in the total order $\sigma$. If $|R|=1$, let $\hg'$ be the subgraph of \hg\ induced by $\{\xx,\yy\}$, otherwise, $|R|>1$ and let $\hg'$ be the subgraph of $\hg$ induced by the vertices in $\cup\{ \rf{u}: u \in R\sm v \}$. By induction, we can untangle $G[V(\hg')]$ to obtain a \pg\ $\overline{\hg'}$ that satisfies the inductive hypothesis and the conditions of the lemma.  It remains to place $v$ and the vertices of $\rf{v}$ that are not yet placed. As described above, these vertices form a path $P$ that goes from some vertex $\xx'$ of $\hg'$ to $v$ to some vertex $\yy'$ of $\hg'$. 

The conditions of the lemma specify the location of $v$. In particular, $v$ is on the \y-axis, with its \y-coordinate equal to its \y-coordinate in $G$. The inductive hypothesis guarantees that the vertex $v$ and any point sufficiently close to $v$ can see\footnote{Given a geometric graph, we say that a point $p$ in the plane \emph{sees} a point $q$, if $\sego{p}{q}$ does not intersect the graph.} all vertices of the outer face of $\overline{\hg'}$. Finally, we note that, if $|R|>1$, then directly below $v$, on the \y-axis, is a vertex $u\in R$. The fact that $u$ is on the \y-axis and that the outer face of $\overline{\hg'}$ is strictly \x-monotone implies that the \x-coordinate of $\xx'$ is less than 0 and that the \x-coordinate of $\yy'$ is greater than 0.  (For the special case when $\xx'=\xx$ and/or $\yy'=\yy$, the above statement is still true.)

Next we place the interior vertices of $P$ to obtain the \pg\
$\overline{\hg}$. To do this, we draw a unit circle $c$, containing $v$, whose center is on the \y-axis and below $v$.  We place all interior vertices of $P$ on $c$ and sufficiently close to $v$ so that:
\begin{enumerate}
\item[$(1)$]  the path on the outer face of $\overline{\hg}$ from \xx\ to \yy\ not containing \xx\yy\ is strictly \x-monotone, 
\item[$(2)$] all interior vertices of $P$ see all other vertices of $P$ in $\overline{\hg}$, 
\item[$(3)$] all interior vertices of $P$ see all vertices on the outer face of
$\overline{\hg'}$ between $\xx'$ and $\yy'$, and
\item[$(4)$] the path on the outer face of $\overline{\hg}$ from \xx\ to \yy\ not containing \xx\yy\ is $\varepsilon$-ray-monotone from $v$ for some
$\varepsilon<\varepsilon'$.

\end{enumerate}
That the first condition can be achieved follows from the fact that
$\xx'$ and $\yy'$ are to the left and right, respectively, of the \y-axis.
That the second condition can be achieved follows from the fact that
we are placing the interior vertices of $P$ on a convex curve (a circle) as
close to $v$ as necessary.  The third condition can be achieved since the upper chain of $\overline{\hg'}$ is $\varepsilon$-ray-monotone from $u$ and hence also from $v$.  That the fourth condition can be achieved follows from the 
definition of $\varepsilon$-ray-monotonicity and the first condition.

Consider the path in $\overline{\hg'}$ from \xx\ to \yy\ not containing \xx\yy\ along the outer face of $\overline{\hg'}$. This path is comprised of the same vertices and edges as a directed path from \xx\ to \yy\ in \fg. Thus, by \lemref{weakly}, the outer face of $\overline{\hg'}$ has no outer chords in $\overline{\hg}$. Therefore, an edge of  $\overline{\hg}$ that is not an edge of $\overline{\hg'}$ is either an edge on $P$, or it is an edge accounted for in Conditions $(2)$ or $(3)$ above. Thus $\overline{\hg}$ is crossing-free. The vertices in $R$ are all on the \y-axis and all have the same \y-coordinates in $G$ as in $\overline{\hg}$. Conditions $(1)$ (and $(4)$) imply that the path between \xx\ and \yy\ on the outer face of $\overline{\hg}$ is strictly \x-monotone.  It remains to show that the internal faces of $\overline{\hg}$ are star-shaped.  The only new faces in $\overline{\hg}$
not present in $\overline{\hg'}$ are the faces having interior vertices of $P$ on their boundary.  However, Conditions $(2)$ and $(3)$ above imply that each
such face is star-shaped from some interior vertex of $P$.  This completes
the proof of the lemma.
\end{proof}

\section{Trees -- upper bound}\seclabel{trees}

In this section we prove the following theorem.

\begin{thm}\thmlabel{trees-lb}
For every positive number $n$ such that $\sqrt{n}$ is an integer, there exists a geometric forest (of stars) $G$ on $n$ vertices, such that $\mv{G}=3(\sqrt{n}-1)$. That is, $G$ cannot be untangled while keeping less than $3(\sqrt{n}-1)$ vertices fixed, and $G$ can be untangled while keeping exactly that many  vertices fixed. 
\end{thm}

\begin{proof}
We first define $G$. A \emph{$k$-star} is a rooted tree on $k+1$ vertices one of which is the root and the rest of the vertices are leaves adjacent to that root. $G$ is a forest on $n$ vertices comprised of trees, $T_i$, $1\leq i\leq \sqrt{n}$, where each $T_i$ is a $(\sqrt{n}-1)$-star. All the vertices of $G$ lie on the \x-axis. For each $i$, the vertices of $T_i$ have the following \x-coordinates $i, i+\sqrt{n}, \dots, i+\sqrt{n}(\sqrt{n}-1)$ where the vertex with the maximum \x-coordinate is the root of $T_i$. This completes the description of $G$.

\noindent{\em Upper bound:}  

We first prove that $\mv{G}\leq 3\sqrt{n}-3$; that is, we prove that $G$ cannot be untangled while keeping more than $3\sqrt{n}-3$ vertices fixed. Let $H$ be an untangling of $G$ with $\mv{G}$ vertices fixed. Let \fl\ denote the number of fixed leaves and \fr\ the number of fixed roots. Let $\fr'$ denote the number of fixed roots that are adjacent to a fixed leaf. Given the ordering of the vertices of $G$ on the \x-axis, it is clear that $\fr'\leq 1$.

Partition the set of free roots into two sets. Let $A$ be the set containing the free roots that are on or above the \x-axis in $H$. Let $B$ be the set containing the free roots that are strictly below the \x-axis in $H$. Our reason for this non-symmetric definition of $A$ and $B$ is to avoid double counting, and not because free roots on the \x-axis have any special meaning. The total number of roots of $G$ is $|A|+|B|+ \fr$.

Suppose that the number of fixed leaves with a neighbour (i.e., a parent) in $A$ is at most $\sqrt{n}-2+|A|$, and similarly for the number of fixed leaves with a neighbour in $B$. As noted above, at most one fixed leaf can be adjacent to a fixed root, thus $\fl\leq 2\sqrt{n}-4 +|A|+|B| + \fr'$. Since $\mv{G}= \fl+\fr$, we get $\mv{G}\leq 2\sqrt{n}-4 +|A|+|B|+ \fr'+ \fr$. Having $|A|+|B|+\fr=\sqrt{n}$ further implies that $\mv{G}\leq 3\sqrt{n}-4 + \fr'$. Since $\fr'\leq 1$, we get the desired upper bound. 

Thus to complete the proof of the upper bound it remains to prove that the number of fixed leaves with a neighbour in $A$ is at most $\sqrt{n}-2+|A|$. The proof below has no special case for the free roots that are on the \x-axis, so the proof for the number of fixed leaves with a neighbour in $B$ is analogous.

Partition the leaves of $G$ into a set of blocks $\{P_j\,:\, 1\leq j\leq \sqrt{n}-1\}$, such that $P_1$ contains the first $\sqrt{n}$ leaves on the \x-axis, $P_2$ the next $\sqrt{n}$ leaves, and so on. More formally, $P_j$ contains all the leaves  with \x-coordinate in the range  $[1+(j-1)\sqrt{n}, j\sqrt{n}]$. 
%
%
Note that each block contains exactly one leaf from each star of $G$. There are $\sqrt{n}-1$ blocks, each containing $\sqrt{n}$ vertices.

Define an auxiliary graph $Q$ with vertex set $V(Q)=A\cup \{p_j\,:\, 1\leq j\leq \sqrt{n}-1\}$, where $vp_j\in E(Q)$ precisely if $v$ is a vertex of $A$ and $v$ has a fixed neighbour in block $P_j$. Thus $Q$ is a bipartite graph, where one bipartition is precisely the set $A$. Note that $|V(Q)|= |A|+\sqrt{n}-1$. Since each vertex of $A$ has exactly one neighbour in each block, the number of fixed leaves whose parents are in  $A$ is precisely $|E(Q)|$. We now show that $Q$ has no cycles. That will complete the proof of the upper bound since in that case $|E(Q)|\leq |V(Q)|-1=|A|+\sqrt{n}-2$.

Assume for the sake of contradiction that $Q$ has a cycle. Let $C$ be a shortest cycle in $Q$. Every second vertex of $C$ is a vertex of $A$. The remaining vertices of $C$ correspond to blocks of leaves. 
Let $C_H$ be the subset of $V(H)$ containing all the roots in $V(C)\cap A$ and for each of those roots, $C_H$ also contains all its fixed leaves contained in blocks $P_j$ for which $p_j$ is in $C$.


Consider the geometric graph $H[C_H]$. The fact that $C$ is a (shortest) cycle and that each vertex in $A$ has exactly one leaf in each block, implies that $H[C_H]$ is a geometric forest of $2$-stars, where the vertices in $V(C)\cap A$ have degree $2$ in $H[C_H]$  and each block $P_j$ such that $p_j\in V(C)\sm A$ has precisely two fixed leaves in $H[C_H]$.


%

Since $H$ is crossing-free, so is $H[C_H]$. Furthermore, since all the roots in $C_H$ are on or above the \x-axis and all the leaves of $C_H$ are on the \x-axis, $H[C_H]$ is fully contained in a closed half-plane determined by the \x-axis. We now show that $H[C_H]$ cannot be crossing-free, which will provide the desired contradiction. $H[C_H]$ is a crossing-free geometric forest of $2$-stars. We first expand $H[C_H]$ into a crossing-free geometric cycle by adding some segments to it, as follows. Consider blocks that contain a leaf of $H[C_H]$. Each such block $P_j$ contains exactly two leaves of $H[C_H]$, denoted by $j_1$ and $j_2$ (see \figref{blocks2}). We claim that $\sego{j_1}{j_2}\cap H[C_H]=\emptyset$. There is no edge of $H[C_H]$ that properly crosses $\sego{j_1}{j_2}$, since $H[C_H]$ is fully contained in a closed half-plane determined by the \x-axis. Therefore, $\sego{j_1}{j_2}\cap H[C_H]$ can be non-empty only if there is an edge of $H[C_H]$ fully contained in $\sego{j_1}{j_2}$. That implies that there is a root of $H[C_H]$ that is located on the \x-axis between $j_1$ and $j_2$. That however is impossible, since one of the two edges of $H[C_H]$ incident to that root would contain $j_1$ or $j_2$ in its interior. This observation implies that $H[C_H]$ can be extend into a \emph{crossing-free} geometric cycle $R$ by adding the appropriate line segments into each block that contains a leaf of $H[C_H]$.
\Figure{blocks2}{\includegraphics{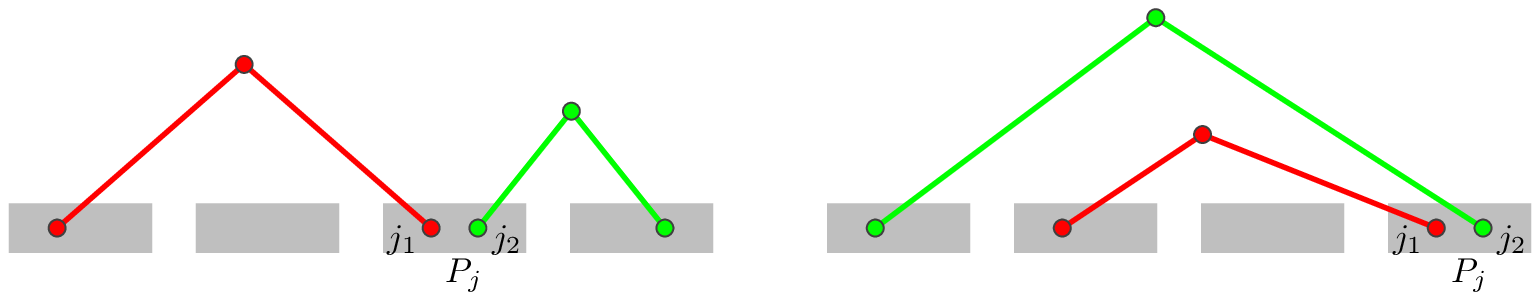}
}{Two $2$-stars (one depicted in green and the other in red) with leaves in a common block.}

Let $v$ be, among all the roots in $C_H$, the one with the smallest index; that is, there is no other root $w\in C_H$ where $v\in T_i$ and $w\in T_j$ and $j<i$.
Vertex $v$ has two neighbours (fixed leaves) in $H[C_H]$,  $s_1\in P_s$ and $t_1\in P_t$ (see \figref{blocks}). Vertex $s_1$ has two neighbours in $R$. One is $v$, and the other is a vertex (fixed leaf) $s_2\in P_s$. Similarly, $t_1$ is adjacent in $R$ to $v$ and to a vertex (fixed leaf) $t_2\in P_t$. Therefore, $R$ contains two vertex disjoint paths: $R_1$, between $s_1$ and $t_1$, and $R_2$, between $s_2$ and $t_2$. Since $v$ belongs to the smallest indexed tree, the ordering of their endpoints on the \x-axis is $s_1 < s_2 <t_1<t_2$. With such ordering of endpoints and since $R$ is fully contained in the closed half-plane above the \x-axis, it is impossible to draw $R_1$ and $R_2$ without crossings (since $R_1$ separates the closed half-plane above the \x-axis into two components, one containing $s_2$ and one containing $t_2$). That is the desired contradiction.
\Figure{blocks}{\includegraphics{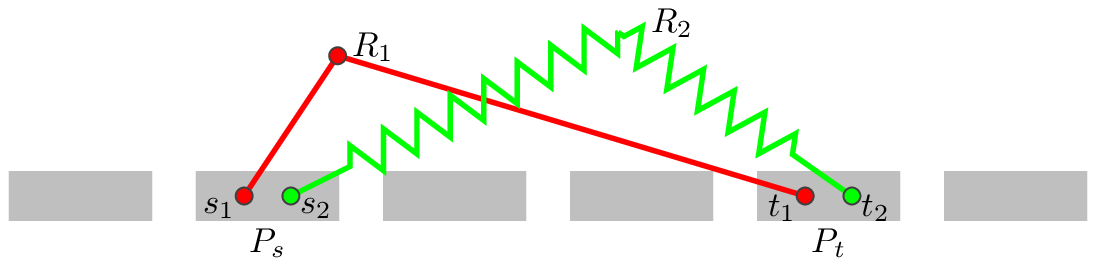}
}{Illustration for the proof of the upper bound of \thmref{trees-lb}.}


%

\noindent{\em Lower bound:} 

We now prove that $\mv{G}\geq 3\sqrt{n}-3$, that is, we prove that $G$ can be untangled while keeping $3\sqrt{n}-3$ vertices fixed. Keep the followings vertices of $G$ fixed:

$(1)$ all the leaves of $T_1$ and $T_2$, and\\
$~~~~~~~~~~~~~~$$(2)$ all the vertices in the block $P_{\sqrt{n}-1}$, and\\
$~~~~~~~~~~~~~~$$(3)$ the root of $T_{\sqrt{n}}$.

Move the root of $T_1$ to the half-plane above the \x-axis and move the root of $T_2$ to the half-plane below the \x-axis. For all $3 \leq i\leq \sqrt{n}-1$, move all the free vertices of $T_i$ to a very small disk centered at the fixed leaf of $T_i$. Move all the free leaves of $T_{\sqrt{n}}$ to a small disk centered at the root of $T_{\sqrt{n}}$. Clearly, this can be done such that the resulting geometric forest $H$ is crossing-free, as illustrated in \figref{trees-lower}. The number of fixed vertices of $H$ is 
$ 2(\sqrt{n}-1) + (\sqrt{n}-2) +1 = 3\sqrt{n} -3$, as claimed.\end{proof}
\Figure{trees-lower}{\includegraphics{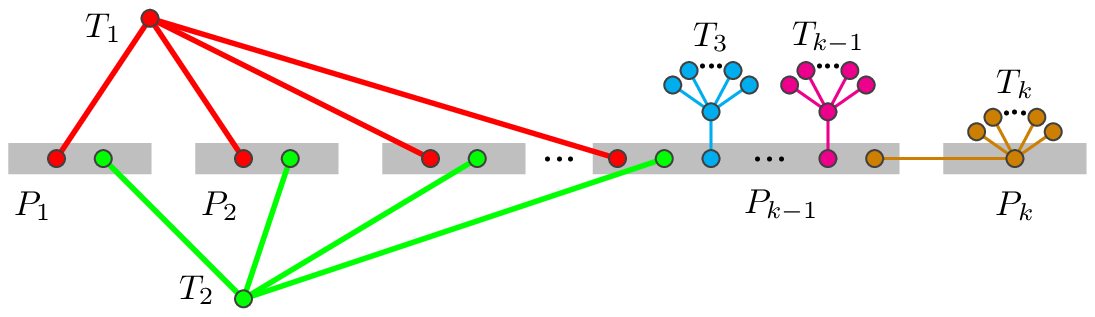}
}{Untangled forest $G$ with $3\sqrt{n}-3$ vertices fixed ($k=\sqrt{n}$).}

\section{Conclusions}

Polynomial bounds are now known for untangling all classes of planar graphs. Tight bounds (up to a constant) are known for untangling trees and outerplanar graphs. The gap remains open for untangling geometric cycles where the best known lower and upper bounds are  $\sqrt{n}$ and $\Oh{(n\log n)^{2/3}}$, and geometric planar graphs where the best known lower and upper bounds are $\Omega(n^{1/4})$ and $\Oh{\sqrt{n}}$.


\section*{Acknowledgements}

This research was initiated at the Bellairs Workshop on Computational
Geometry for Geometric Reconfigurations, February 1st to 9th, 2007.  The
authors are grateful to Godfried Toussaint for organizing the workshop
and to the other workshop participants for providing a stimulating
working environment.

\bibliographystyle{myBibliographyStyle}
\bibliography{paper,2trees}
\end{document}